\documentclass[
    ,final            
  ]
  {aipproc}

\layoutstyle{8x11single}

\begin{document}

\title{Low-energy magnetic radiation: deviations from GOE\footnote{Part of the material has been published in \cite{Schwengner14}}}
\classification{}
\keywords      {Shell Model, M1 strength function, magnetic rotation, Gaussian Orthogonal Ensemble}

\author{S. Frauendorf}{
  address={University Notre Dame, IN 46557, USA}
}

\author{R. Schwengner}{
  address={IRP, HZDR, 01328 Dresden, Germany}
}

\author{K. Wimmer}{
  address={Central Michigan University,  Mt. Pleasant, MI 48859, USA }
}

\begin{abstract}
 A pronounced spike at low energy in the strength function for magnetic radiation (LEMAR) is found by means of Shell Model calculations,
 which explains the experimentally observed enhancement  of the dipole strength. 
 LEMAR originates from statistical low-energy M1-transitions between many excited complex states.
  Re-coupling of the proton and neutron high-j orbitals generates the strong magnetic radiation. LEMAR is 
 closely related to Magnetic Rotation. LEMAR is predicted for nuclides participating in the r-process of element synthesis
 and is expected to change the reaction rates. An exponential decrease of the strength function and a power law for the
 size distribution of the $B(M1)$ values are found, which strongly deviate from the ones of the GOE of random matrices,  which is 
 commonly used to represent complex compound states. 
    \end{abstract}

\maketitle


\section{Low-energy Enhancement of Gamma Radiation}

\begin{figure}[h]
\begin{minipage}{0.48\textwidth} 
\includegraphics[width=\textwidth]{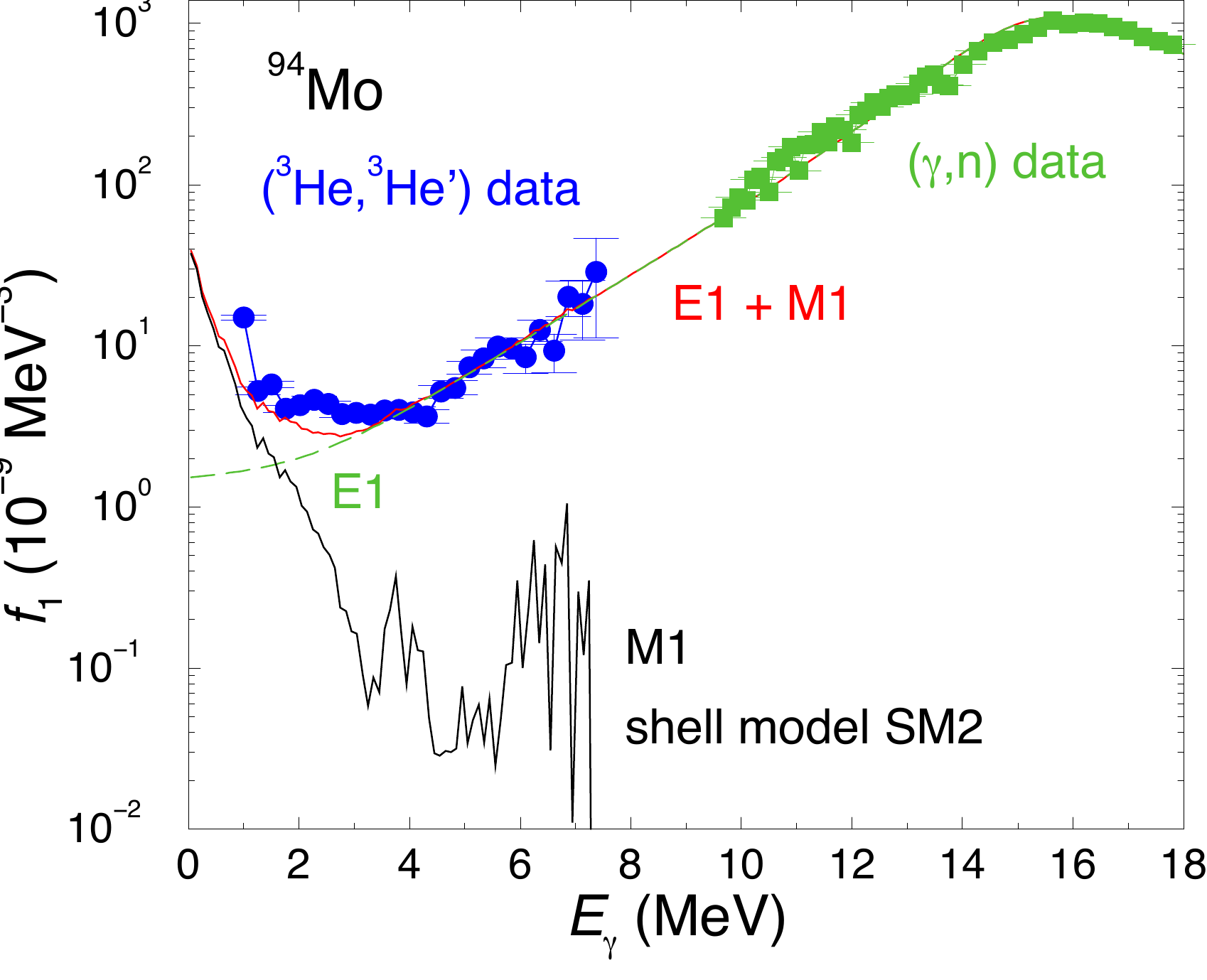}
\end{minipage}
\begin{minipage}{0.48\textwidth}
\vspace*{-0.5cm}
\includegraphics[width=1.17\textwidth]{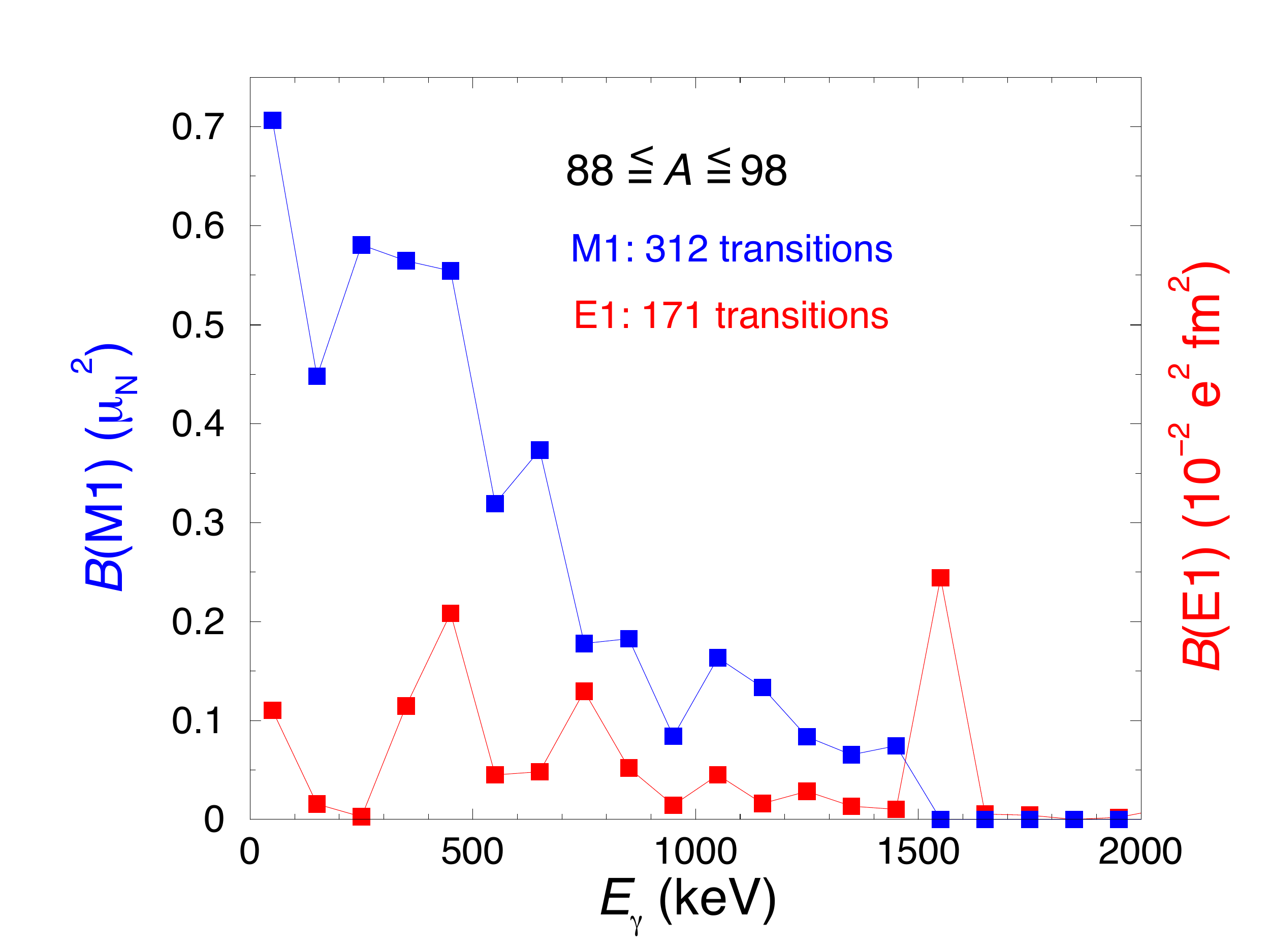}
\end{minipage}
\caption{(Color online) {\bf Left panel:}  Strength functions for $^{94}$Mo 
deduced from ($^3$He,$^3$He') (blue circles) and $(\gamma,n)$ (green squares)
experiments, the $M1$ strength function from the present Shell Model
calculations (black solid line), $E1$ strength according to the GLO analytical
expression (green dashed line), and the total ($E1 + M1$) dipole strength
function (red line). 
{\bf Right panel: } Average  
reduced transition probabilities of  discrete dipole transitions in all nuclides with $88\leq A \leq 98$
as reported in ENSDF \cite{ENSDF}. The transitions are sorted into bins of 100 keV of the transition energy.
Reprinted with permission from Ref. \cite{Schwengner14}.\label{fig:94MoEgf1}}
\end{figure}

  Photonuclear reactions and the inverse radiative-capture reactions between
nuclear states in the region of high excitation energy and large level density
are of considerable interest in many
applications. Radiative neutron capture, for example, plays a central role in
the synthesis of the elements in various stellar environments, for next-generation nuclear technologies, and
as the transmutation of long-lived nuclear waste. 
A critical input to
calculations of the reaction rates is the average strength of the cascade of 
$\gamma$-transitions  de-exciting the nucleus,
which is described by photon strength function.
 Modifications of its low-energy 
part can cause drastic changes in the
abundances of elements produced via neutron capture in the r-process occurring
in violent stellar events \cite{lar10}.
Such  an increase of the
dipole strength function below 3 MeV toward low $\gamma$-ray energy has recently been
observed  in  nuclides in the mass range from $A \approx$ 40 to 100. In particular, this
low-energy enhancement of the strength function was deduced from experiments
using ($^3$He,$^3$He') reactions on various Mo isotopes \cite{gut05}. 
The ($^3$He,$^3$He') data for $^{94}$Mo are shown in
Fig.~\ref{fig:94MoEgf1} (left). Around \mbox{1 MeV}, the experimental strength function (blue) is about a factor of 10 
larger than expected for a damped Giant Dipole Resonance shown by the dashed green curve, which 
is calculated by the standard GLO expression commonly used for describing the strength of 
electric dipole (E1) radiation  in this energy region.

The enhancement is not observed in the inverse process of absorbing $\gamma$-quanta
by nuclei in the ground state. 
Only few discrete lines are found within the interval of the first 4 MeV  \cite{Pietralla99}. 
The enhancement in the de-excitation cascade must be related to the complex structure of the highly
excited states among which the transitions occur. Fig. \ref{fig:94MoEgf1} (right) shows the summed reduced 
probabilities of all  discrete transitions reported for the nuclides  with $88\leq A \leq 98$ depending on their 
transition energy. The reduced probabilities of the magnetic transitions $B(M1)$ clearly increase toward zero transition energy,
whereas no such tendency is seen for reduced probabilities $B(E1)$ for the electric transitions. Based on this observation we 
conjectured that the enhancement seen in experiments like the one in Fig. \ref{fig:94MoEgf1} (left) is caused by $M1$ transitions between
high-lying states.  To study this conjecture, we carried out Shell Model calculations for 
the nuclides $^{94,95,96}$Mo and $^{90}$Zr, for which the enhancement has been observed in experiment.

\newpage

\section{Shell Model Calculations}

The Shell Model calculations were performed within two model spaces. The first
one (SM1) included the active proton orbits
$\pi(0f_{5/2}, 1p_{3/2}, 1p_{1/2}, 0g_{9/2})$ and the neutron orbits
$\nu(1p_{1/2}, 0g_{9/2}, 1d_{5/2})$ relative to a  $^{66}$Ni core. 
The second one (SM2) included the same proton orbits, but the active neutron
orbits $\nu(0g_{9/2}, 1d_{5/2}, 0g_{7/2})$ relative to a $^{68}$Ni core. 
We discuss only the results for (SM2), because the ones for (SM1) are very similar. 
Details about the set of empirical matrix elements for the effective
interaction and of the single particle energies in this model space  are
 given in  Refs. \cite{Schwengner14,sch09}. Our previous
studies 
of  nuclei with $N = 46 - 54$ demonstrated that the present Shell Model  
very well accounts for the experimental energies and transition probabilities 
of discrete states near the yrast line (see Ref.~\cite{sch09} and
references therein). For calculating the  reduced transition probabilities $B(M1)$  effective
$g$-factors of $g^{\rm eff}_s = 0.7  g^{\rm free}_s$ have been applied.

To make the calculations feasible truncations of the occupation numbers were
applied. In SM2, up to two protons could
be lifted from the $1p_{1/2}$ orbit to the $0g_{9/2}$ orbit. In $^{94,95}$Mo, one
neutron from the $0g_{9/2}$ orbit could be excited to either the $1d_{5/2}$ or
the $0g_{7/2}$ orbit, and one from the $1d_{5/2}$ to the $0g_{7/2}$ orbit. In
$^{90}$Zr, one neutron from the $0g_{9/2}$ orbit may be excited to the 
$1d_{5/2}$ orbit and one from the $1d_{5/2}$ orbit to the $0g_{7/2}$ orbit,
or one neutron from the $0g_{9/2}$ orbit and one from the $1d_{5/2}$ orbit
may be excited to the $0g_{7/2}$ orbit. 

The calculations included states with spins from $J$ = 0 to 6 for $^{90}$Zr and
$^{94}$Mo and from $J$ = 1/2 to 13/2 for $^{95}$Mo. For each spin the lowest 40
states were calculated. The reduced transition probabilities $B(M1)$  were
calculated for all transitions from initial to final states with energies
$E_f < E_i$ and spins $J_f = J_i, J_i \pm 1$. For the minimum and maximum
$J_i$, the cases $J_f = J_i - 1$ and $J_f = J_i + 1$, respectively, were
excluded. This resulted in more than 14000 $M1$ transitions for each parity
$\pi = +$ and $\pi = -$, which were sorted into 100 keV bins according to
their transition energy $E_\gamma = E_i - E_f$. The average $B(M1)$ value for
one energy bin was obtained as the sum of all $B(M1)$ values divided by the
number of transitions within this bin. The results for 
$^{94}$Mo are shown in Fig.~\ref{fig:94MoEgM1} (left). 
They look quite similar for the other nuclides studied. 
Clearly there is a spike at zero energy that extends to about 2 MeV, which we
call  {\bf Low-Energy MAgnetic Radiation (LEMAR).} 

The inset of Fig.~\ref{fig:94MoEgM1}(left) demonstrates that,
up to 2 MeV, the LEMAR spike of $\overline{B}(M1,E_\gamma)$ is well
approximated by the exponential function
\begin{equation}
\overline{B}(M1,E_\gamma) = B_0 \exp{(-E_\gamma/T_B)},
\end{equation}
 with $B_{0} = \overline{B}(M1,0)$ and $T_B$ being constants. This is the case for all studied cases.
 For the respective
parities $\pi = +, -$ we find
for $^{90}$Zr: $B_0$ = (0.36, 0.58) $\mu^2_N$ and $T_B$ = (0.33, 0.29) MeV, 
for $^{94}$Mo: $B_0$ = (0.32, 0.16) $\mu^2_N$ and $T_B$ = (0.35, 0.51) MeV, 
for $^{95}$Mo: $B_0$ = (0.23, 0.12) $\mu^2_N$ and $T_B$ = (0.39, 0.58) MeV, and
for $^{96}$Mo: $B_0$ = (0.20, 0.13) $\mu^2_N$ and $T_B$ = (0.41, 0.50) MeV.

\begin{figure}[h]
\begin{minipage}{0.48\textwidth} 
\hspace*{-0.59cm}
\includegraphics[width=1.16\textwidth]{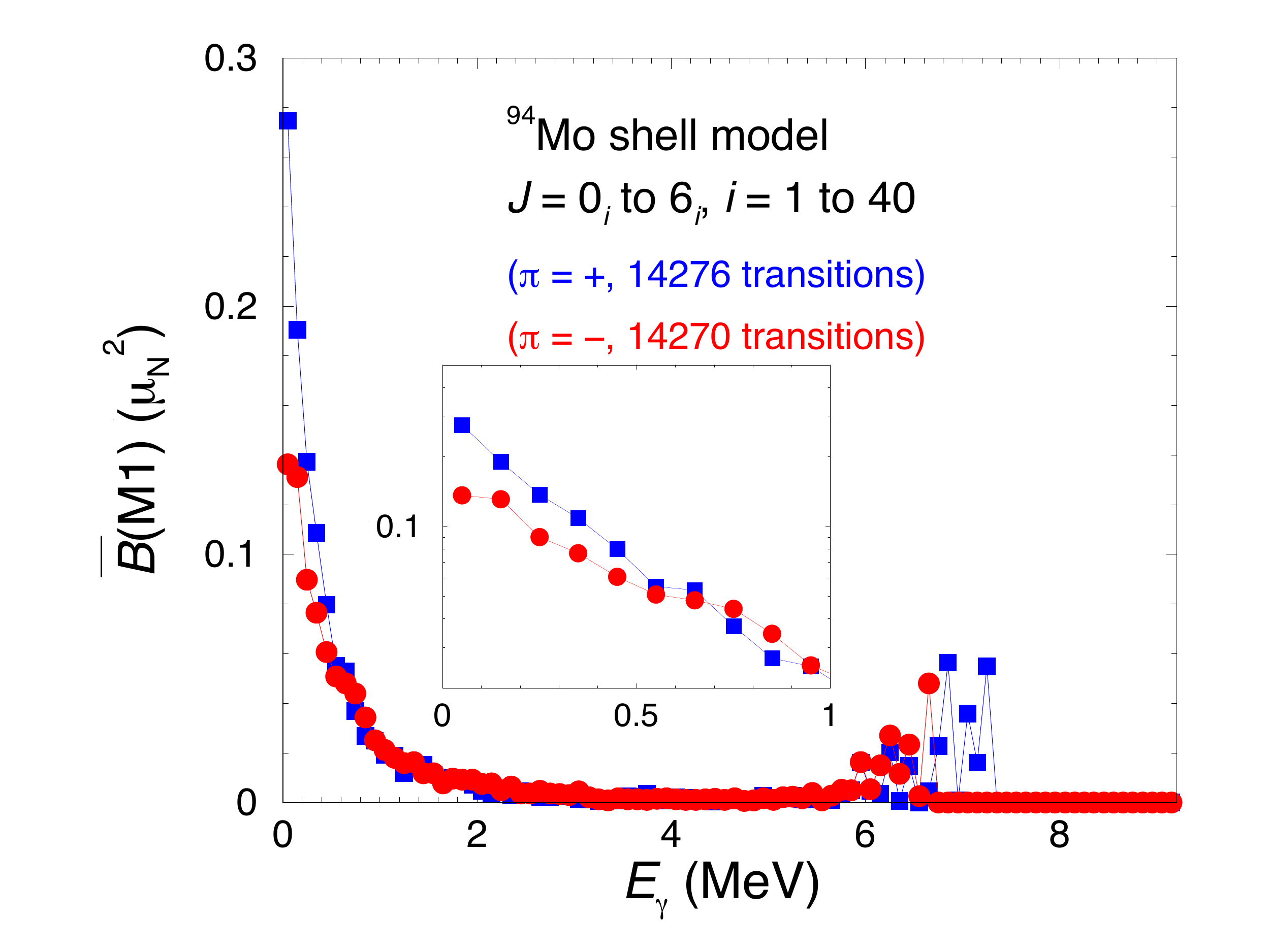}
\end{minipage}
\begin{minipage}{0.48\textwidth}
\vspace*{-0.3cm}
\includegraphics[width=1.22\textwidth]{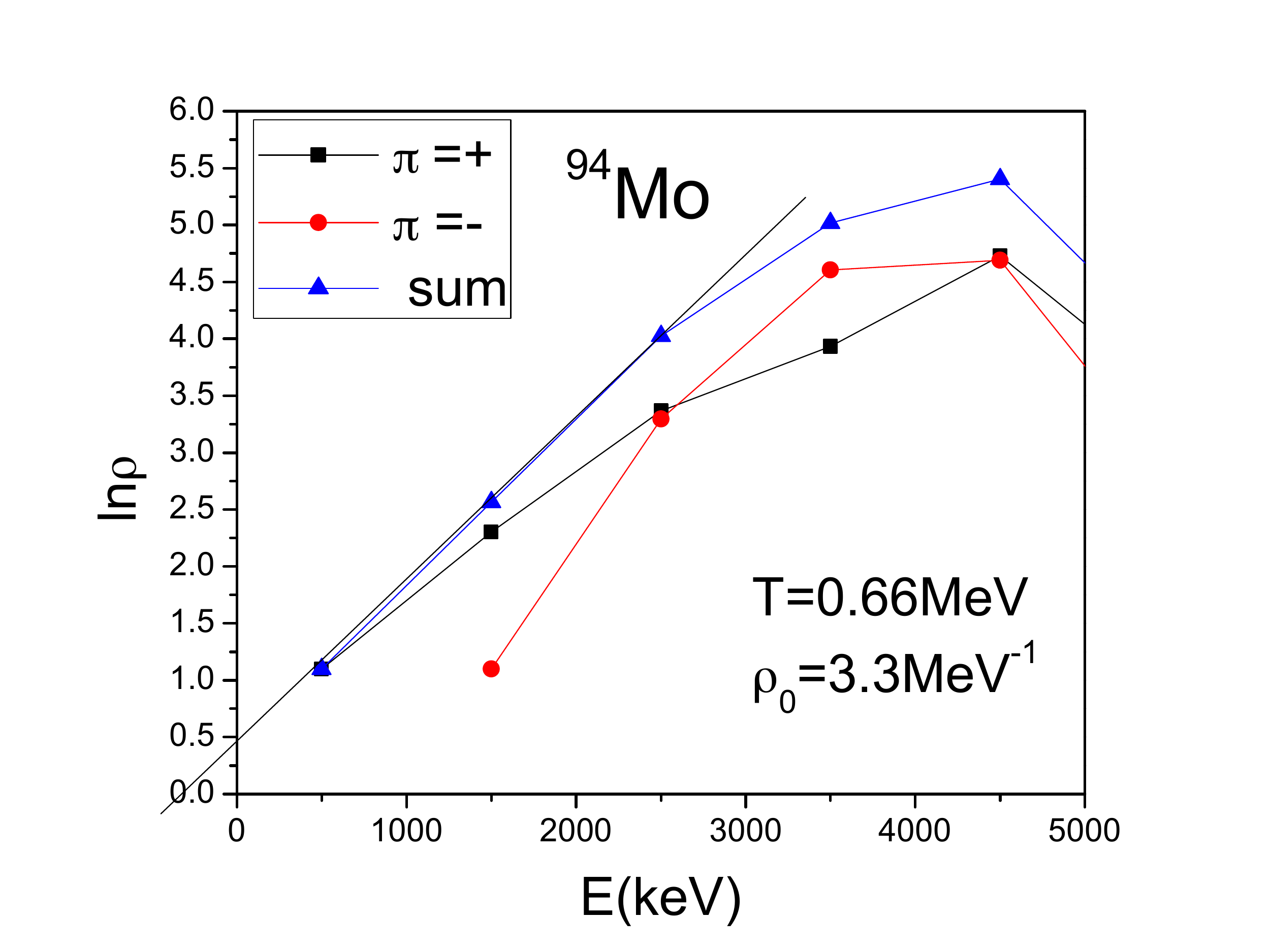}
\end{minipage}
	\caption{(Color online) \label{fig:94MoEgM1} {\bf Left panel:} Average $B(M1)$ values in 100 keV
bins of the transition energy calculated for positive-parity (blue squares) and
negative-parity (red circles) states in $^{94}$Mo. The inset shows the
low-energy part in logarithmic scale. Reprinted with permission from Ref. \cite{Schwengner14}.
{\bf Right panel:} Level density of $^{94}$Mo calculated as the number of levels within bins of 1 MeV. \label{fig:94MoEgM1}}

\end{figure}
\begin{figure}[h]
\begin{minipage}{0.48\textwidth} 
\hspace*{-0.59cm}
\includegraphics[width=1.15\textwidth]{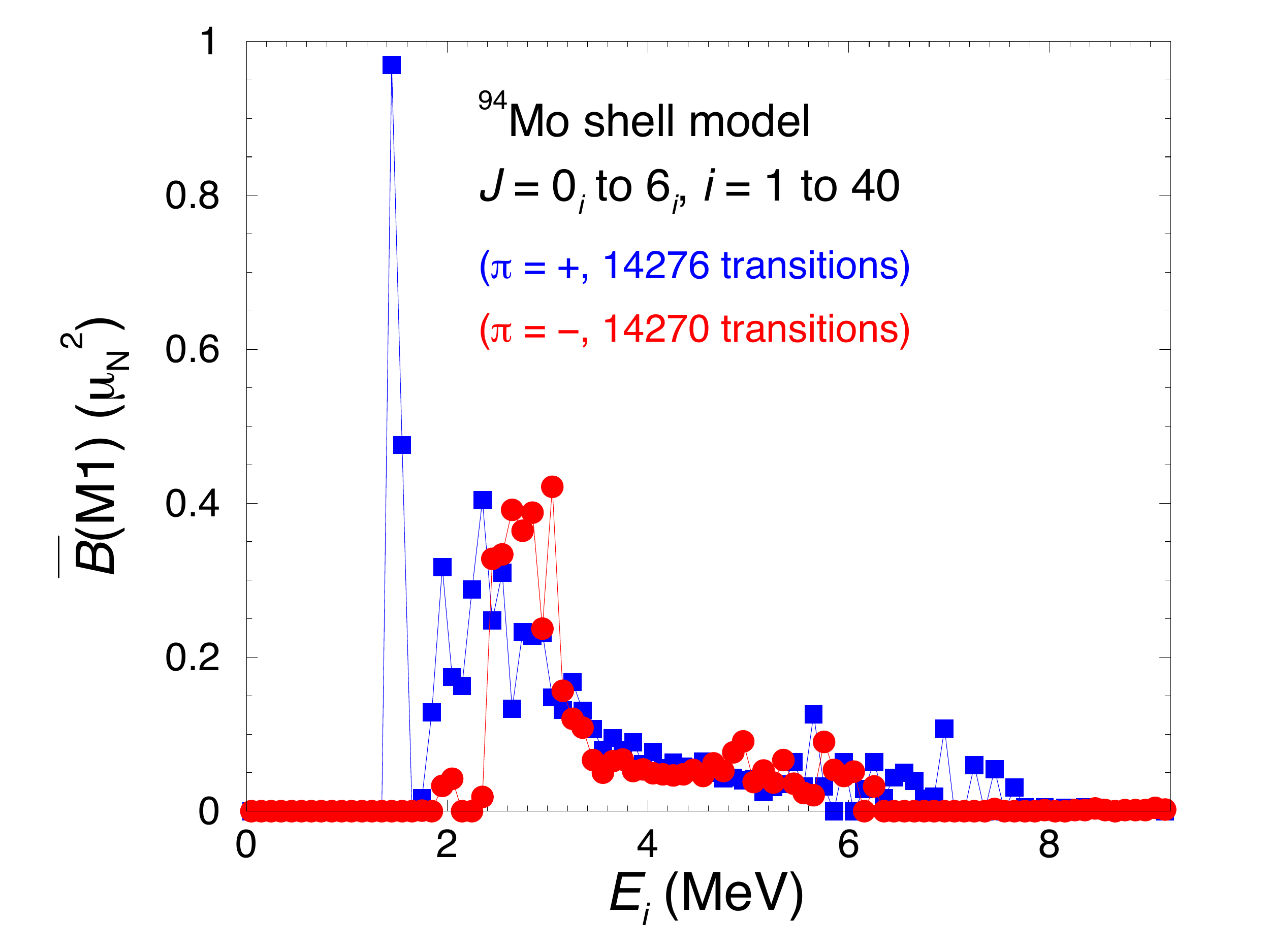}
\end{minipage}
\begin{minipage}{0.48\textwidth}
\includegraphics[width=1.15\textwidth]{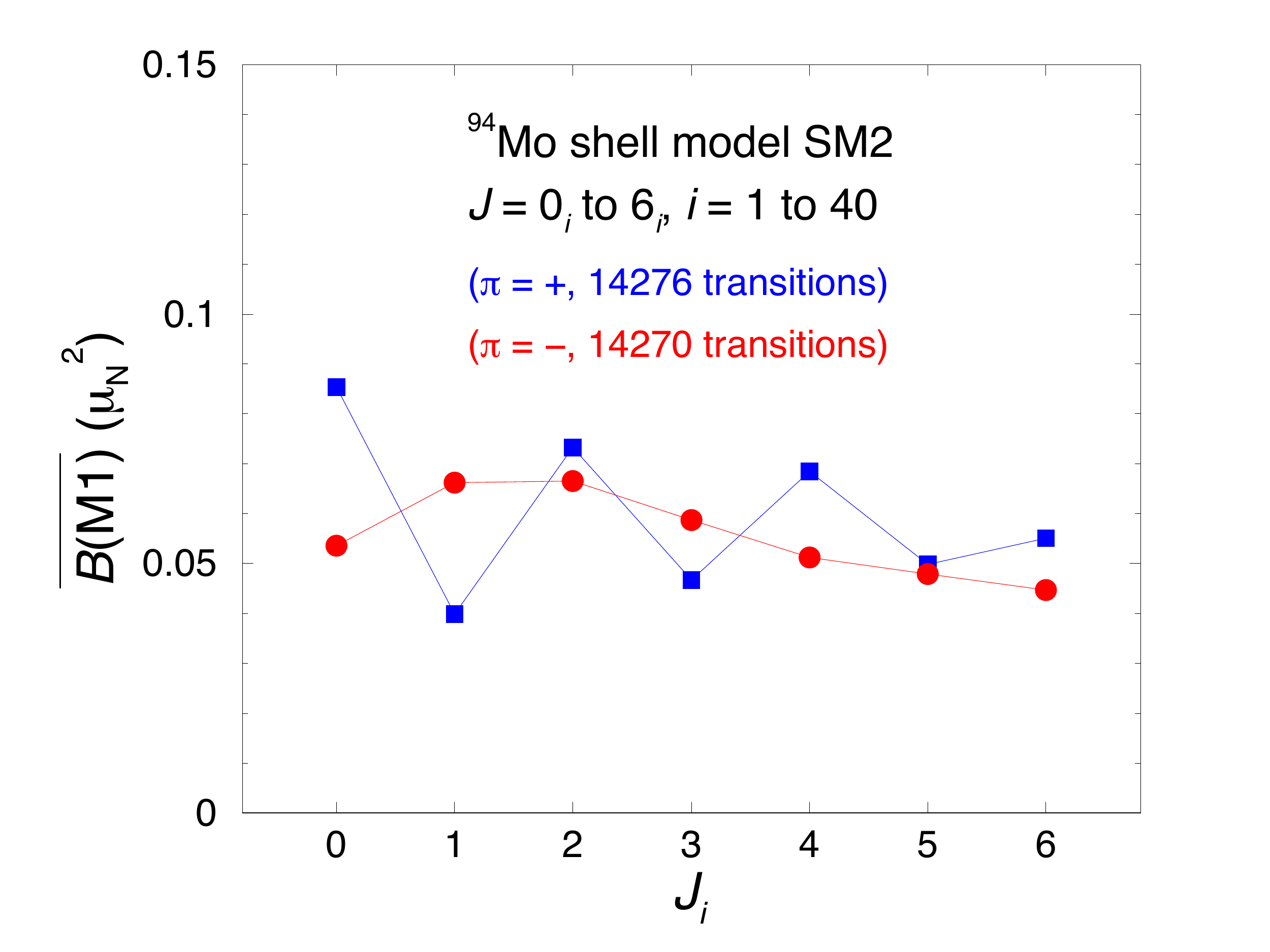}
\caption{(Color online) {\bf Left panel:}  Average $B(M1)$ values in 100 keV
bins of the excitation energy calculated for positive-parity (blue squares) and
negative-parity (red circles) states in $^{94}$Mo. {\bf Right panel: } Average of the $B(M1)$ values for transitions 
originating from the  lowest 40 initial state of given angular momentum $J_i$.
Reprinted with permission from Ref. \cite{Schwengner14}.\label{fig:94MoExM1}}
\end{minipage}
\end{figure}

The  exponential dependence on the transition energy is retained by
the $M1$ strength functions, which are defined by the relation 
\begin{equation}
f_{M1}(E_\gamma)
= 16\pi/9 (\hbar c)^{-3}\overline{B}(M1,E_\gamma)\rho(E_i),
\end{equation}
where the level density at the initial state  $\rho(E_i)$ is obtained from the Shell Model calculations.
 The level densities
$\rho(E_i,\pi)$ were determined by counting the calculated levels within energy
intervals of 1 MeV for the two parities separately. 
These combinatorial level densities were used in calculating the strength functions
 by means of Eq. (2). Fig. \ref{fig:94MoEgM1} (right) shows the level density for $^{94}$Mo for both parities.
As seen  in Fig. \ref{fig:94MoEgf1} (left),  there is a pronounced enhancement below 2 MeV, which is
well described by the exponential function
\begin{equation}
f_{M1}(E_\gamma) = f_0 \exp{(-E_\gamma/T_f)}.
\end{equation}
For $^{90}$Zr, $^{94}$Mo, $^{95}$Mo, and $^{96}$Mo, the parameters are
$f_0$ = (34, 37, 39, 55) $\times$ 10$^{-9}$ MeV$^{-3}$ and 
$T_f$ = (0.50, 0.50, 0.51, 0.48) MeV, respectively.
The calculated M1- enhancement is consistent with the experiment, 
which however did not determine wether the radiation is electric or magnetic.

To find out which states generate strong $M1$ transitions, the average
$\bar{B}(M1)$ values for $^{94}$Mo are plotted as a function of the energy of
the initial states in Fig.~\ref{fig:94MoExM1} (left). 
The $\overline{B}(M1)$ distributions versus $E_i$ in $^{95}$Mo and $^{96}$Mo look
similar to the ones in $^{94}$Mo, but are shifted to somewhat lower excitation
energy. In $^{90}$Zr, the distributions start at about 3 MeV and continue to 10 MeV.
Fig. \ref{fig:94MoExM1} (right) shows that the transitions  
from initial states with different angular momentum within the studied range of $0 \leq J\leq 6$
have about the same  probability. 
The distributions for the other nuclides are similar. 

To summarize:
{\bf
\begin{itemize}
\item LEMAR is generated by a huge number of weak low-energy  $M1$ transitions.
\item They originate from high-lying states.
\item They add up to strong $M1$ radiation.
\item LEMAR accounts for the observed low-energy enhancement of the strength function.
\end{itemize}
}

\section{Origin of the M1-strength}

\begin{figure}[h]
\begin{minipage}{0.48\textwidth} 
\hspace*{-0.65cm}
\includegraphics[width=1.17\textwidth]{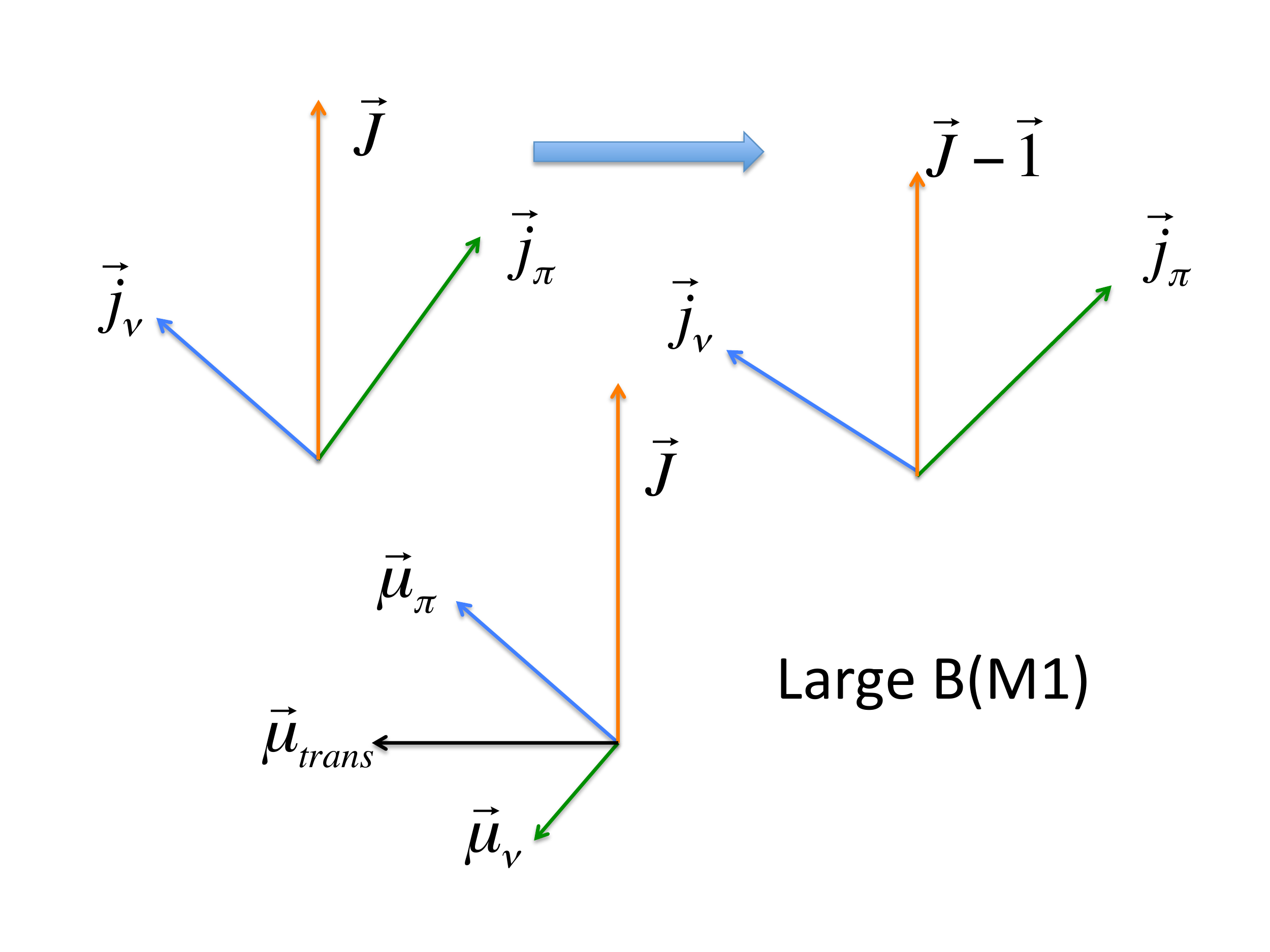}
\end{minipage}
\begin{minipage}{0.48\textwidth}
\includegraphics[width=1.2\textwidth]{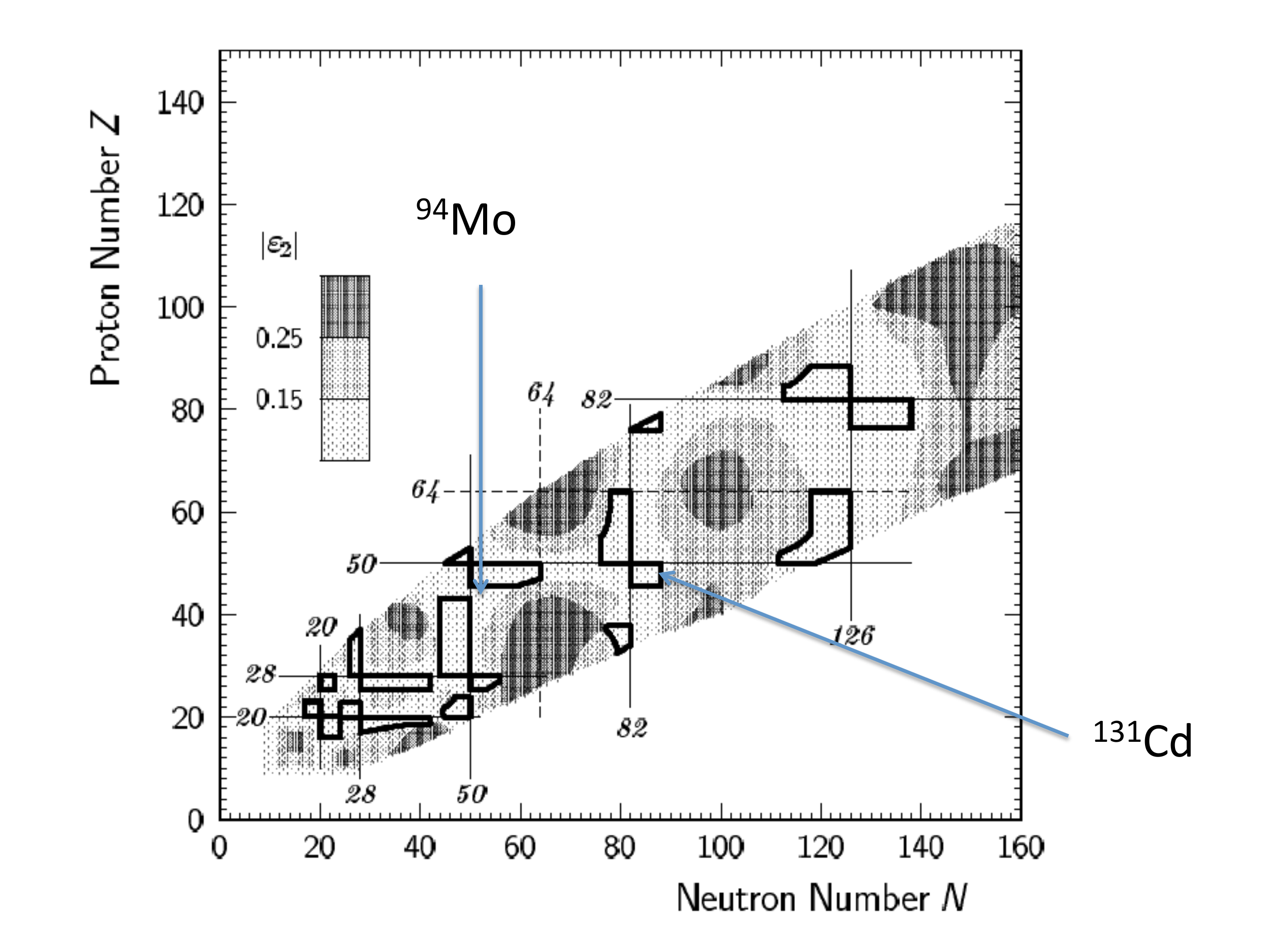}
\caption{(Color online) {\bf Left panel:}Magnetic Rotation  as an example for large
values of $B(M1)$ between two states generated by coupling the protons in one configuration
($\vec j_\pi$) and neutron holes in another configuration ($\vec j_\nu$) to total spin $J$ and $J-1$.
Semiclassically,  $B(M1)\approx 3/8\pi \mu^2_{trans}$, because the transverse magnetic moment $\vec \mu_{trans}$
rotates about the axis $\vec J$ generating strong magnetic radiation.   
Along a rotational band the angular momenta of the protons and neutron holes gradually align
against  the repulsive residual interaction between the orbitals, which results in the regular spacing $\hbar \omega$
 between the band members. The $M1$ radiation is very strong, because the $j=l+1/2$ orbitals have  
 large  absolute values of the $g$-factors, and the transverse components of the magnetic
 moments $\vec \mu_\pi$ and $\vec \mu_\nu$ add, because   
  $g_\pi>0$ and $g_\nu<0$.   For details see Ref. \cite{fraRMP}.
   {\bf Right panel:} Regions where Magnetic Rotation is observed or predicted (inside the black boundaries). The conditions for 
appearance are: A combination of high-j proton particle orbitals with  high-j neutron holes (or vice versa)
and small deformation. Reprinted with permission from Ref.  \cite{fraRMP}.\label{fig:MR}}
\end{minipage}
\end{figure}

LEMAR is caused by transitions between
many close-lying states of all considered spins located well above the yrast
line in the transitional region to the quasi-continuum of nuclear states.  
Inspecting the composition of initial and final states, one finds large $B(M1)$ values for transitions
between states that contain a large component (up to about 50\%) of the same
configuration with broken pairs of both protons and neutrons in high-$j$
orbits. The largest $M1$ matrix elements connect configurations with the spins
of high-$j$ protons re-coupled with respect to those of high-$j$ neutrons to
the total spin $J_f = J_i, J_i \pm 1$. The main configurations are
$\pi(0g_{9/2}^2) \nu(1d_{5/2}^2)$, 
$\pi(0g_{9/2}^2) \nu(1d_{5/2}^1 0g_{7/2}^1)$, and 
$\pi(0g_{9/2}^2) \nu(1d_{5/2}^2 0g_{9/2}^{-1} 0g_{7/2}^1)$ for positive-parity
states in $^{94}$Mo. Negative-parity states contain a proton lifted from the
$1p_{1/2}$ to the $0g_{9/2}$ orbit in addition. The orbits in these configurations have large 
$g$-factors with opposite signs for protons and neutrons. Combined with specific
relative phases of the proton and neutron partitions they cause large total
magnetic moments.

For states near the yrast line, the re-coupling of spin  generates
the ``Shears Bands'' manifesting ``Magnetic Rotation'' (MR) \cite{fraRMP}, 
the mechanism of which is described in Fig. \ref{fig:MR} (left). Regular MR bands 
are dominated by the configurations of $j=l+1/2$ orbitals, which generate easily angular momentum and $M1$ radiation.   
MR was observed in the mass 90 region \cite{sch99}. 
 When   less favorable orbtitals are an essential component of the configuration, the sequences become less regular
(see e.g \cite{frau97}).   Typical transition energies are about 0.5 MeV both for the regular and irregular sequences.

The residual interaction between the valence particles and holes generates an energy difference between the states 
related to each other by recouping the angular momenta, which were degenerated without it. These energetic splittings enable 
transitions between the states by emitting an $M1$ $\gamma$-quant. In this sense, it is the residual interaction that generates the radiation.
MR bands are located close to the yrast line.  
Accordingly, the configurations are rather pure, and the transition energies increase with angular momentum in a regular way.  
The high-lying states that generate the LEMAR
 are composed of a strong mix of configurations. The complex mixing changes the residual interaction   between
the states. As a  consequence,  the distance between the states becomes randomized. As discussed below, the mean transition energy
of the LEMAR strength function is 0.5 MeV, which is the typical transition energy for  MR bands.

To summarize:
{\bf
\begin{itemize}
\item LEMAR consists of transitions between states related by angular momentum recoupling
of the same high-j proton and neutron configurations.  
\item Transition energies and probabilities are randomized.
\item LEMAR is closely related to Magnetic Rotation.
\end{itemize}
}
\section{Regions where LEMAR is expected}

\begin{figure}[h]
\includegraphics[width=0.92\textwidth]{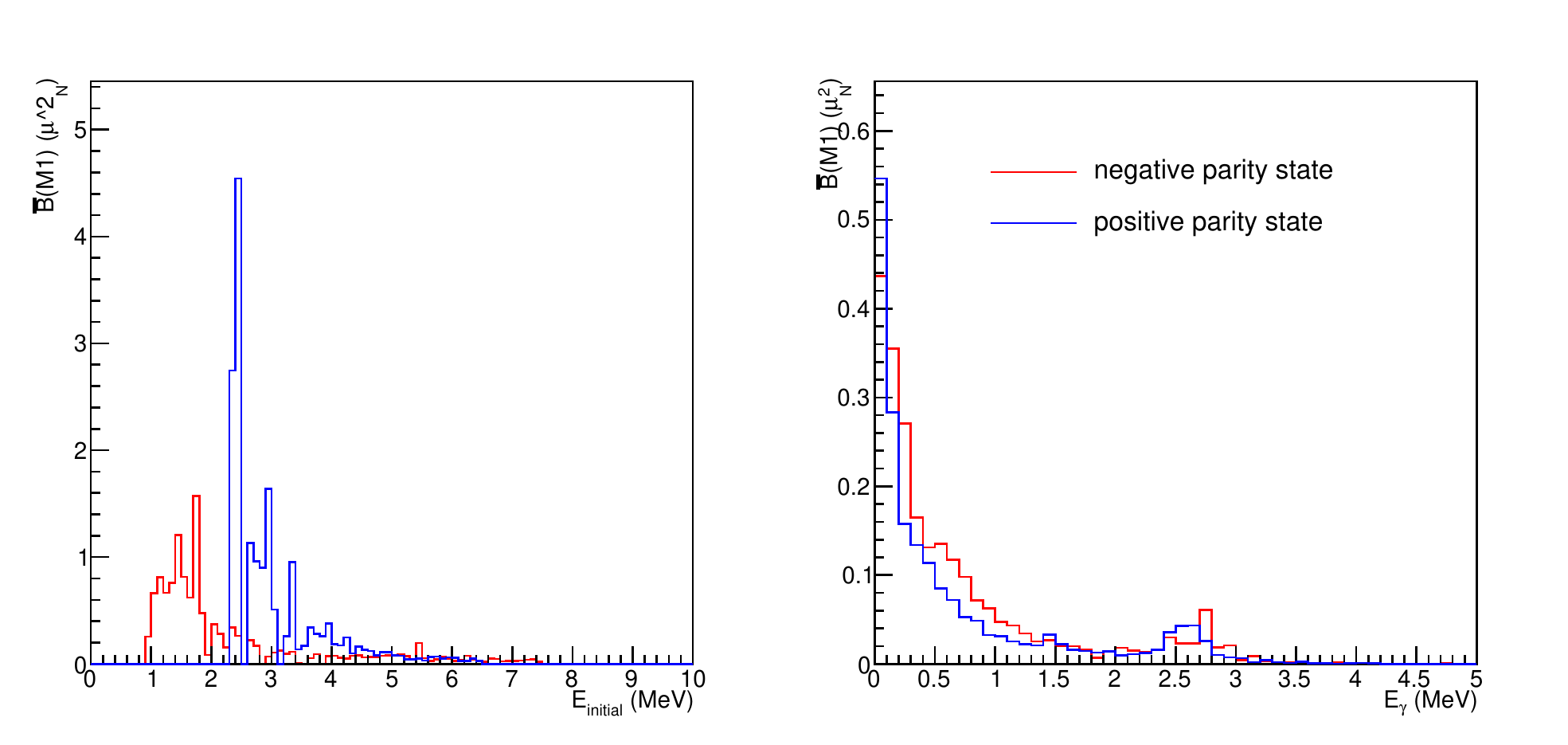}
\caption{\label{fig:Cd131} Right panel:  Average $B(M1)$ values in 100 keV
bins of the transition energy calculated for  $^{131}$Cd. Left panel:  Average $B(M1)$ values in 100 keV
bins of the excitation energy calculated for $^{131}$Cd.
}
\end{figure} 

The close relation between MR and LEMAR suggests that both phenomena appear in the same nuclei.
 The regions in the nuclear
chart, where MR is expected, are delineated in Fig. \ref{fig:MR} (right).
 In fact, $^{90}$Zr and the Mo isotopes discussed in the
present work as well as the Fe, Ni, and Cd isotopes, for which the low-energy
enhancement was observed, belong to these regions. On
the other hand, $^{117}$Sn, $^{158}$Gd, and the Th,
Pa isotopes, for which no low-energy enhancement was observed, lie
outside these regions (see Ref. \cite{Schwengner14} for references).

According to Fig. \ref{fig:MR} (right), MR is predicted for the region with proton number below $Z=50$ and 
neutron number above $N=82$. These nuclei play a key role in the r-process of element synthesis in violent stellar events.
For this reason we studied $^{131}$Cd, which is a waiting point in the reaction chain. The Shell Model calculation 
was performed within the model space of the  $1p_{3/2}$, $0f_{5/2}$, $1_{p1/2}$, $0g_{9/2}$   proton holes and
 $0h_{9/2}$, $1f_{7/2}$, $1f_{5/2}$, $2p_{3/2}$, $2p_{1/2}$ neutrons particles using a 
  G-matrix derived from the CD-Bonn NN interaction. Fig. \ref{fig:Cd131} shows the results, which are quite similar 
  to the ones for the stable Mo isotopes.  As demonstrated in Ref. \cite{lar10}, a low-energy enhancement of the dipole strength function comparable 
  to the one observed in stable nuclei will substantially change the abundances of the elements synthesized in the r-process.    
  Calculations of element abundances using  $\gamma$-strength functions that include the LEMAR spike are on the way. 
  
To summarize:
{\bf
\begin{itemize}
\item LEMAR is expected where MR appears. 
\item Nuclides with observed or missing low-energy enhancement in the $\gamma$-strength
 function correspond to the ones where LEMAR is expected.
 \item LEMAR is predicted is for $^{131}$Cd.
\item Strong modifications of the r-process rates are expected.
\end{itemize}
}

\section{Statistical Properties of the Transitions}
As the LEMAR is generated by a huge number of weak transitions 
between complex  states, it is natural to study  their statistical
characteristics.  The study is still on the way, and we report only tentative results
obtained so far.

\begin{figure}[h]
\begin{minipage}{0.48\textwidth} 
\hspace*{-0.59cm}
\includegraphics[width=1.05\textwidth]{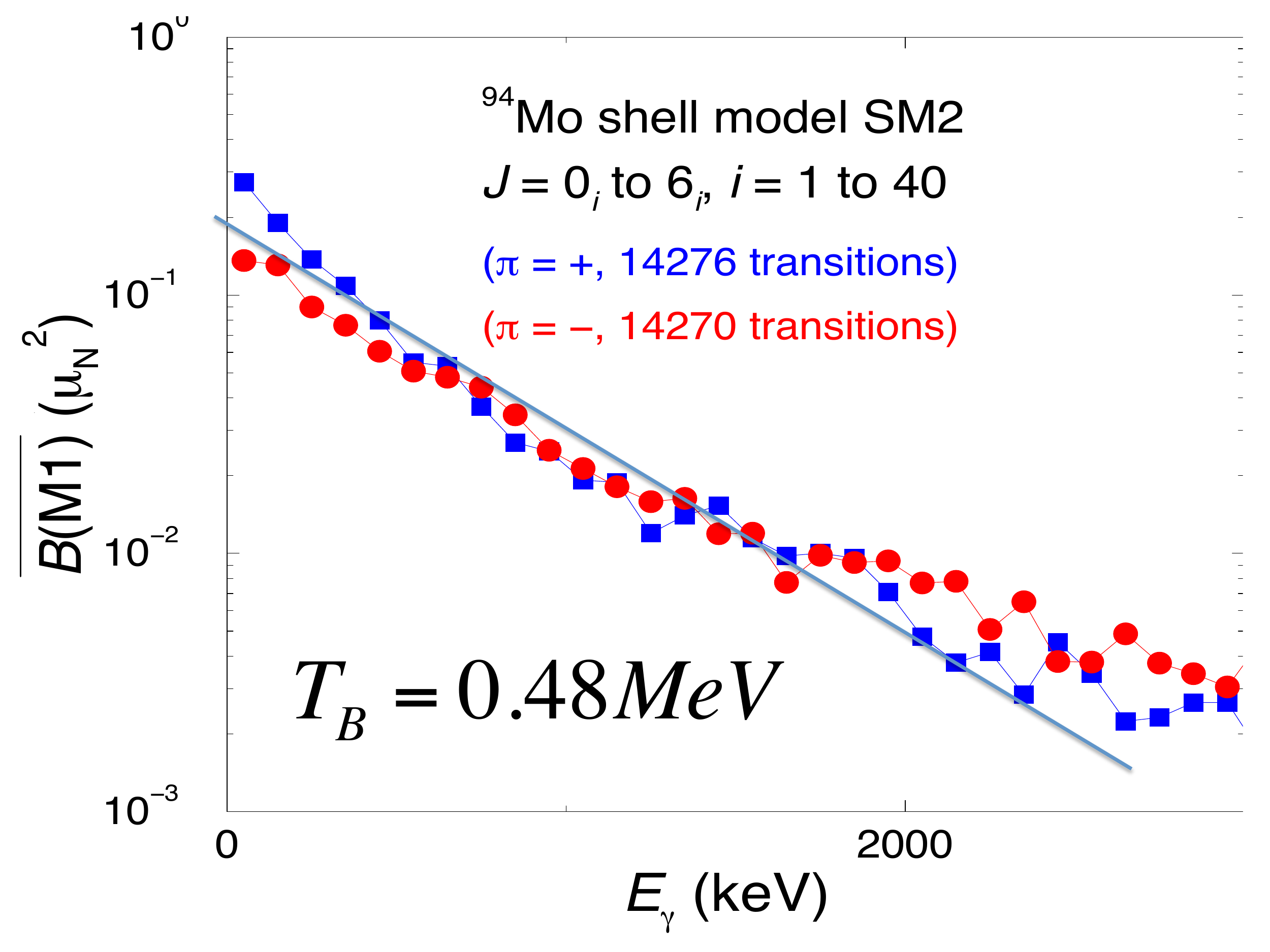}
\end{minipage}
\begin{minipage}{0.48\textwidth}
\vspace*{0.4cm}
\includegraphics[width=1.05\textwidth]{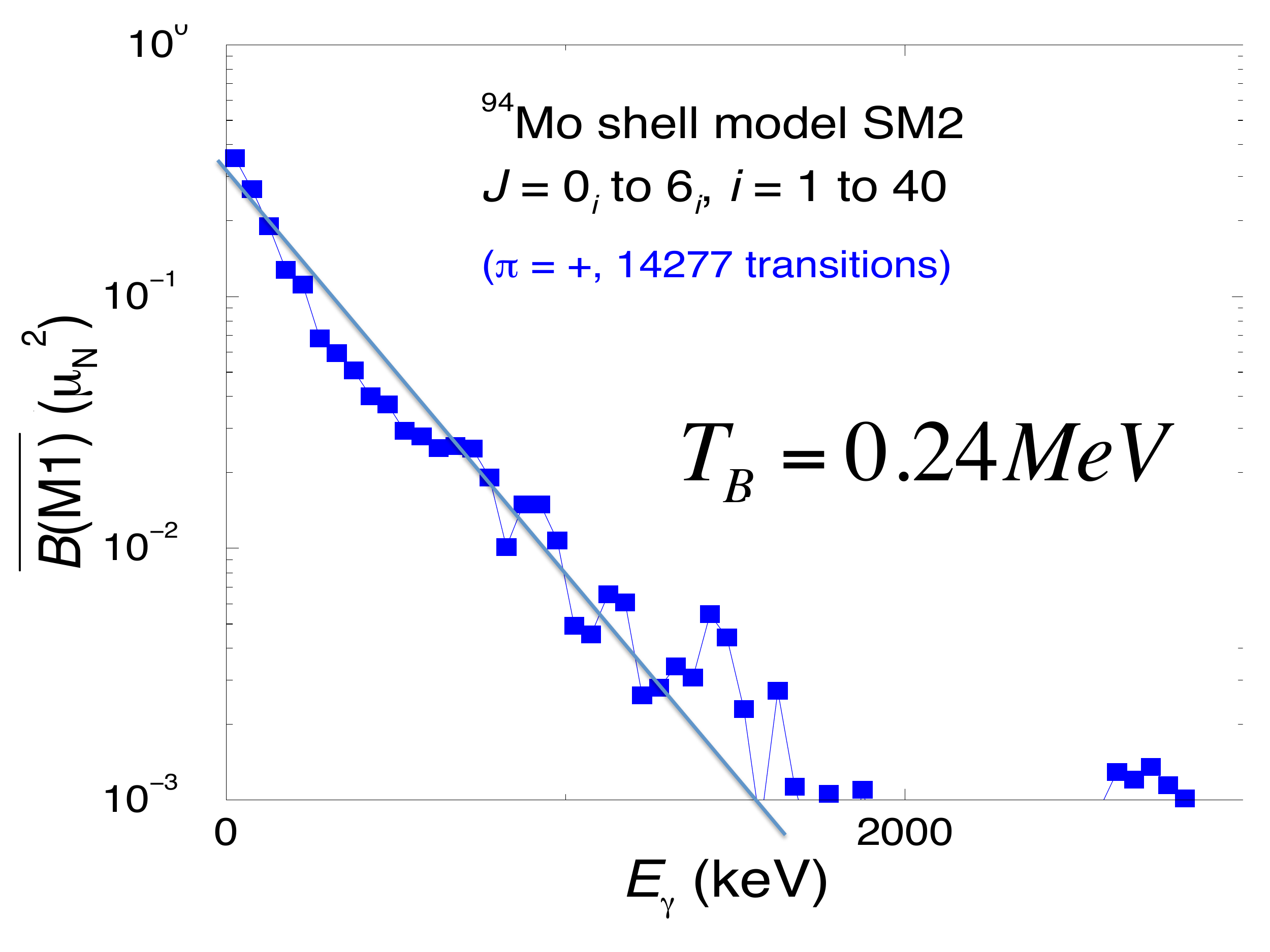}
\caption{(Color online)  Average $B(M1)$ values in 100 keV
bins of the transition energy calculated for positive-parity (blue squares) and
negative-parity (red circles) states in $^{94}$Mo. {\bf Left panel:}  full interaction,  {\bf Right panel: }
half interaction\label{fig:94MoBM1full}}
\end{minipage}
\end{figure}

As already pointed out and illustrated by the left panel of Fig. \ref{fig:94MoBM1full}, the average reduced transition probabilities $B(M1)$ decrease exponentially with the
 energy difference between initial and final states of the transitions. The decrease is determined by the 
 parameter $T_B$ in Eq. (1), which scatters around 0.5 MeV. The strength functions decrease exponentially 
 as well, with the characteristic parameter in Eq. (3)  $T_f\approx 0.5$ MeV.  The mean value of the transition energy for 
 the strength function is $T_f\approx 0.5$ MeV, which is the typical transition energy for MR.
Such pronounced exponential dependence
  is unusual. More common are a weak dependence on the transition energy or a Lorentzian   resonance 
 caused by some doorway state (e. g. the Giant Dipole Resonance in the $E1$ strength function).

Fig. \ref{fig:94MoEgM1} (right) shows that the total level density $\rho(E_i)$ is well reproduced by the
constant-temperature expression
\begin{equation}
 \rho(E_i) = \rho_0 \exp{(E_i/T_\rho)}
 \end{equation}
  as long
as $E_i <$ 3 MeV. For higher energies the combinatorial level density deviates
from this expression and eventually decreases with excitation energy, which is
obviously due to missing levels at high energy in the present configuration
space. The level densities for $^{95,96}$Mo are similar. 
 From a fit to the combinatorial values in the range $E_i < 2$ MeV we
found for $(\rho_0, T_\rho)$ in (MeV$^{-1}$, MeV) the values of (1.37, 0.67),
(1.90, 0.54), and (1.25, 0.58) for $^{94}$Mo, $^{95}$Mo, and $^{95}$Mo,
respectively. The level density in the semi-magic $^{90}$Zr shows a more
complicated energy dependence. 
  It is noticed that the micro canonical temperature $T_\rho \sim 0.6$ MeV, 
is close to $T_M\sim  0.5$ MeV, which  determines  the exponential  
decrease of the  average $B(M1)$ values with the transition energy.
 
\begin{figure}[h]
\begin{minipage}{0.48\textwidth} 
\hspace*{-0.59cm}
\includegraphics[width=1.05\textwidth]{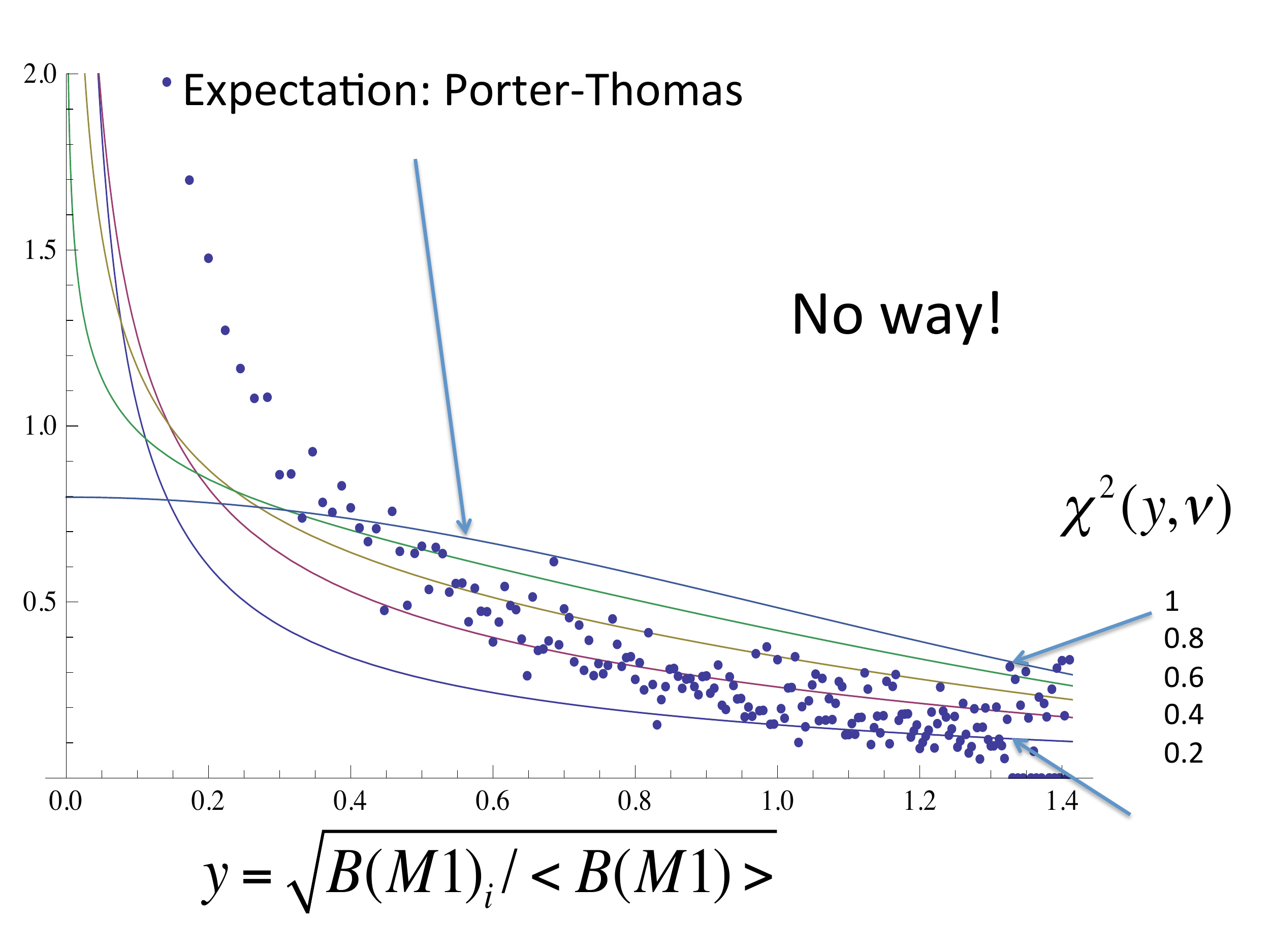}
\end{minipage}
\begin{minipage}{0.48\textwidth}
\vspace*{0.4cm}
\includegraphics[width=1.05\textwidth]{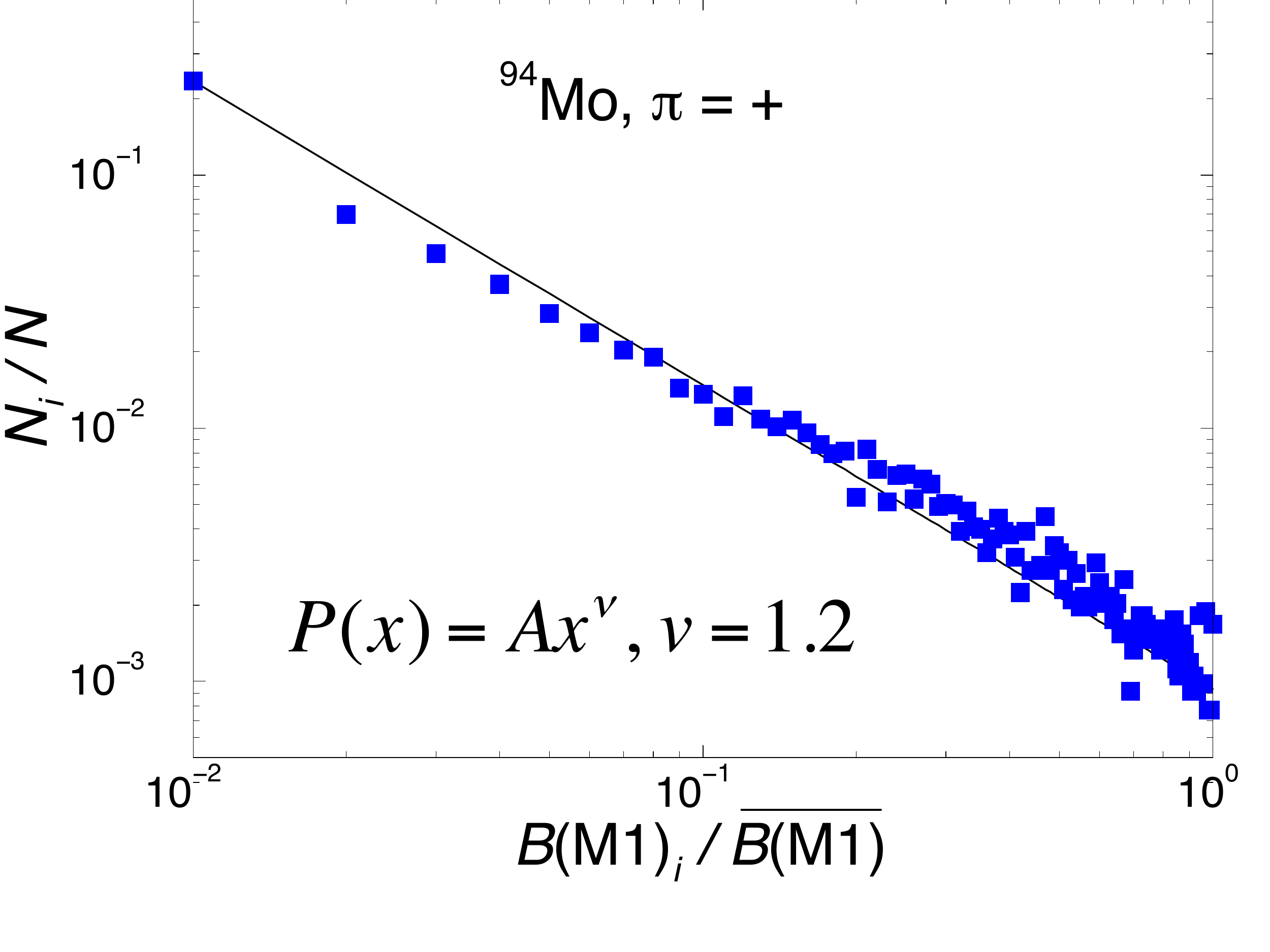}
\caption{(Color online)   {\bf Left panel:}   Probability distribution of the $B(M1)$ values in $^{94}$Mo for positive parity states
compared with $\chi^2$ distributions
of different index $\nu$, where $\nu=1$ is the Porter-Thomas distribution. {\bf Right panel: }
Probability distribution of the $B(M1)$ values in $^{94}$Mo for positive parity states
compared with a power law distribution (straight line).\label{fig:94MoBM1}\label{fig:NoPT} }
\end{minipage}
\end{figure}

The size distribution of the $B(M1)$ values is shown in Fig. \ref{fig:NoPT}.
Based on   Shell Model  \cite{ham02} and experimental \cite{shr00} 
studies of $\gamma$- transitions between complex 
excited states, we expected a Porter-Thomas-like distribution. As seen in the left panel,
our distribution turned out to be very different.  Also the more general $\chi^2(y,\nu)$
distributions for various indices $\nu$ do not account for our distribution.
Quite surprisingly, we found that the distribution follows a power law
\begin{equation}
 P(y) = Ay^\nu ~~~ y=B(M1)/\overline{B(M1)},
 \end{equation}
which is illustrated by the right panel.
Here, $\overline{B(M1)}$ is the mean value over the complete distribution.
The distributions for  both parities in all three nuclides follow a power law with
the exponents $\nu$ scattering around 1.2.

The Porter-Thomas distribution characterizes the Gaussian Orthogonal Ensemble (GOE) of
random matrices (see Ref. \cite{wei09} for a recent review).  Its appearance is taken as a signature
of onset of many-body chaos in a quantum system, in which time reversal symmetry is conserved. 
The LEMAR distribution of the $B(M1)$ values deviates drastically from the GOE. 
The reason may be the fact that LEMAR is generated by the subset of transitions between
special states that are related by re-coupling of angular momentum.  
These states have the same energy without residual interaction. In contrast,
 the orthogonality 
condition of the GOE requires  that  its random matrices have 
 Gaussian probability distributions for both the diagonal and the non-diagonal matrix elements,
 where the width of the former is twice the width of the latter.  As discussed in Ref. \cite{zel04},
 it may also be that the considered valence space and the sampled range of excitation energies 
 are too small for approaching the chaotic regime. 

The fact that the LEMAR distributions can be described by a simple power law is remarkable in our view.
So far we have not found a conclusive explanation. Power-law distributions are characteristic for scale-free 
systems, as for example the distribution of the number of clicks/site in the internet (Any site can connect with any number of other sites.)
or the  heat capacity near a second order phase transition (There is no scale for the fluctuations of the order parameter.).
LEMAR may be classified as  scale-free as follows. The various configurations that are related by re-coupling of the angular momentum
have all the same energy. The energy difference between the mixed states is generated by their residual interaction, which acts in a random 
way between the complex states. This differs from the conventional situation, where the single particle level spacing   
represents an energy scale for the various configurations to mixed by the residual interaction. 

Statistical self-similarity is another  signature 
of scale-free systems. It means that the statistical characteristics of the system are the same when studied on different  scales
(The length of coast lines is a popular example.). LEMAR seems to exhibit statistical self-similarity. The right panel of Fig. \ref{fig:94MoBM1full} shows the average $B(M1)$
values calculated within the same model space and multiplying each matrix element of the residual interaction by a factor of 0.5 (i. e. changing the scale).  Comparing  with the 
results for the full interaction in in the left panel, it is seen that the distribution 
remains exponential, where the characteristic parameter $T_B=0.24$ MeV is one half of $T_B=0.48$ MeV of the  full calculation.    

Finally, we remark that the exponential growth of the low-energy level density with excitation energy  has been attributed to the 
quenching of the pairing correlations, which corresponds to the phase transition of second order observed in macroscopic
superconductors and supra fluids. Magnetic transitions are known to be sensitive to the presence of pair correlations.
It is possible that the phenomena are related, and the closeness of the parameters $T_\rho \sim T_B \sim 0.5$ MeV is more than accidental.

To summarize:
{\bf
\begin{itemize}
\item The LEMAR strength function decreases exponentially with increasing transition energy.
\item The combinatorial level density increases exponentially with the excitation energy.  
 \item The two parameters that control the respective decay or growth are  $T_B \sim T_\rho\sim 0.5$ MeV. 
\item The size distribution of the reduced transition probabilities is a power law with an exponent of $\sim 1.2$.
\item LEMAR representing a scale-free system might be the reason for these unexpected statistcal properties.
\end{itemize}
}

Our considerations concerning 
 the exponential decrease of the LEMAR strength function and of the power law 
of the $B(M1)$ size are speculative so far. A more profound analysis is on the way, which may provide a 
better understanding of the nature and origin of these unexpected phenomena.

\begin{theacknowledgments}
 Supported by the Grant No. DE-FG02-95ER4093 of the US Department of Energy.
 \end{theacknowledgments}



\bibliographystyle{aipproc}   


\end{document}